\documentclass[preprint]{revtex4}
\usepackage{graphicx}% Include figure files
\usepackage{dcolumn}% Align table columns on decimal point
\usepackage{bm}% bold math
\begin{document}
\newcommand{\vs}{}
\title{Thermodynamics from a scaling Hamiltonian}
 \author{L. A. del Pino$^1$, P. Troncoso$^2$ and S. Curilef$^2$}%\altaffiliation[e-mail:]{scurilef@ucn.cl}
\affiliation{%
$^1$Facultad de Monta\~na, Universidad de Pinar del Rio, Cuba.\\
$^2$Departamento de F\'\i sica, Universidad Cat\'olica del Norte,
Av. Angamos 0610, Antofagasta, Chile.}
\date{\today}
\begin{abstract}
There are problems with defining the thermodynamic limit of systems
with long-range interactions; as a result, the thermodynamic
behavior of these types of systems is anomalous. In the present
work, we review some concepts from both extensive and nonextensive
thermodynamic perspectives. We use a model, whose Hamiltonian takes
into account spins ferromagnetically coupled in a chain via a power
law that decays at large interparticle distance $r$ as
$1/r^{\alpha}$ for $\alpha\geq0$. Here, we review old nonextensive
scaling. In addition, we propose a new Hamiltonian scaled by
$2\frac{(N/2)^{1-\alpha}-1}{1-\alpha}$ that explicitly includes
symmetry of the lattice and dependence on the size, $N$, of the
system. The new approach enabled us to improve upon previous
results. A numerical test is conducted through Monte Carlo
simulations. In the model, periodic boundary conditions are adopted
to eliminate surface effects.
\end{abstract}
%\keywords{Applications of Monte Carlo methods; Thermodynamics; Classical spin models}%Use showkeys class option if keyword
\maketitle \section{Introduction}\vs A good description of magnetic
ordering phases illuminates concepts about the critical behavior and
possible applications of magnetic devices. We know that the magnetic
behavior of systems decreases when the dimensionality of a physical
system decreases. This led to incorrect beliefs that prompted
specialists to lose interest in one-dimensional systems. However,
several theoretical and practical aspects, which appeared in the
physical properties of one-dimensional systems, caused specialists
to reverse those
beliefs.\cite{pre70,mplb17,pra45,prl76,prl86,prb56,prb34,PGaNat416,Zedler,Curilef,prbCannas}

Recently, ferromagnetism in one dimension has been reported in
various systems. Transitions between two different magnetic ordering
phases are obtained using different approaches and considerations,
and we will discuss some of these. First, microscopic anomalies lead
to important modifications in the thermodynamic properties of
systems. As a consequence, it is suggested that anisotropy barriers
contribute to the novel effect.\cite{PGaNat416} Second, the Gibbs
potential of a one-dimensional metal at constant magnetization is
calculated to second order in the screened electron-electron
interaction. At zero temperature, a possible
paramagnetic-ferromagnetic quantum phase transition was found in
one-dimensional metals, which must be first order.\cite{Zedler}
Finally, a special Hamiltonian considers long-range interactions
through a power law that decays at large interparticle distances. It
has been shown that if the range of interactions decreases, the
critical temperature trend disappears, but if the range of
interactions increases, the trend of the critical temperature
approaches the mean field approximation. The crossover between these
two limiting situations is preliminarily discussed.\cite{Curilef}.
That crossover is a consequence of long-range interactions, and we
use it to illustrate the nonextensive perspective of the
thermodynamics.

To describe the behavior of systems with long range interactions,
some scalings approaches~\cite{kac} have been introduced in the
literature. In the present work, we carefully discuss the
thermodynamic behavior of systems with microscopic long-range
interactions, taking into account a method of implementing periodic
boundary conditions (section~\ref{PBC}) via a scaling Hamiltonian
(section~\ref{scal}). Previously, the critical temperature between
two states of different magnetic ordering was obtained in a
spin-$\frac{1}{2}$ Ising linear chain, where the range of
interactions is, at least, comparable to the size of the chain.
However, in order to obtain suitable thermodynamic behavior, in
accordance with impositions from extensivity, we improve the so
called nonextensive scaling (section~\ref{scal}) for a symmetric
one-dimensional lattice.

This paper is organized as follows: In section~\ref{nonext}, we
introduce the nonextensive perspective of thermodynamics. In
section~\ref{PBC}, we explain the method of implementing periodic
boundary conditions to eliminate surface effects. In
section~\ref{scal}, we review the Tsallis scaling to get a formalism
of the thermodynamics from a scaling Hamiltonian and then explain
some important results about the presented formalism. In
section~\ref{conclu} we summarize our work and make some concluding
remarks.
\section{Nonextensive Thermodynamics}\label{nonext}
In this section, we introduce some fundamental facts about the
nonextensive perspective of thermodynamics, which can be easily
illustrated by long-range interaction systems.\vspace{-1cm}
\subsection{Tsallis Scaling:}\vspace{-0.5cm} This method is useful for
scaling thermodynamic quantities that depend strongly on the size of
the system with long-range microscopic interactions. The explicit
form of Tsallis scaling appears by evaluating the internal energy
per particle. We take into account interparticle potentials $v(r)$
with an attractive tail that decays as:
\begin{equation}\label{powerlaw}
  v(r)= \frac{1}{r^\alpha}.
\end{equation}
The original Tsallis scaling, from the asymptotic trend of $U_N$,
for systems in one dimension, is given by:

\begin{equation}
N^* = \left\{\begin{array}{llll}
  &(N^{1-\alpha})/{(1- \alpha)} &\text{for } &0\leq\alpha<1 \\
 &\log(N) &\text{for } &\alpha=1\\
 &1/{(\alpha-1)} &\mbox{for } &\alpha > 1.
\end{array}\right.\label{scaling_b}
\end{equation}
The present scaling has been revised by several authors and applied
to different physical situations. The scaling was expressed in a
most appropriate form for discrete systems\cite{prbCannas} and was
previously written as $2^\alpha N^*$. Indeed, in the present work we
reviewed this kind of scaling and included it as a best approach.
\subsection{Mean Field Approximation} We consider a one dimensional
model of $N$ spins-$\frac{1}{2}$. If the coupling  is long-range,
surface effects appear and become important. This fact requires
working with very large systems. However, surface effects can be
ignored, even in moderately sized systems, by applying periodic
boundary conditions. This is a nontrivial task when long range
interactions are considered. However, it is possible to do in a
suitable form, which is introduced in section~\ref{PBC}. The model
is described by the Hamiltonian:
\begin{equation}
H = -\sum_{i,j =1}^n J(i-j) s_i s_j, \label{Hs}
\end{equation}
where $n$ is the number of particles in every cell, and $s_i= \pm 1
\;\;\;\forall i$. The difference $(i-j)$ represents the distance
between two sites. A non external magnetic field is considered. The
coupling decays as a power law given by:
\begin{equation}\label{Jij}
 J(i-j) = \frac{J}{|i-j|^\alpha},
\end{equation} where $J$ is a
positive parameter which measures the strength of the coupling. In
previous works, these kinds of systems have been discussed by
several
authors.\cite{pre70,mplb17,pra45,prl76,prl86,prb56,prb34,Curilef,prbCannas,Tsallis}
A common behavior has been conjectured~\cite{Tsallis} for generic
systems with arbitrary (long or short)-range interactions. With the
aim of calculating the mean value of the Hamiltonian, we proceed in
the same way as in the Eq.(\ref{powerlaw2}). $\label{H2} \langle H
\rangle = \sum_{i=1}^{N} E(N)s_{i} $, where $E(N) =\frac{1}{2} J
{N}^* \langle s \rangle $; the factor $\frac{1}{2}$ ensures that we
do not count the same pair of spin twice and in this view; ${N}^*$
(defined in the Eq.(\ref{powerlaw2})) represents the {\em effective}
number of the  nearest neighbor spins. The quantity $ \langle s
\rangle$ is the average spin per site. This assumption helps us
recover all the results for systems with long-range microscopic
interactions in a similar manner to the traditional mean field
approximation. This treatment can be extended even if the external
magnetic field is nonzero.

So, the average spin per site of the lattice is given by $\label{sm}
 \langle s \rangle = \tanh\left[\frac{1}{2}\beta {N}^*J \langle s
 \rangle\right].
$ In the absence of an external magnetic field, the magnetization is
zero for a high temperature paramagnetic phase, and it will be
nonzero at lower temperatures where the spins have spontaneously
aligned.
 For the present system, the internal energy is given by
$ U(N) = \frac{1}{2}N {N}^* J \langle s \rangle^{2}. $ Hence, the
internal energy is given by $U(N)=NN^*U_1$. The extensive property
imposes observables to be a linear homogeneous function of $N$ and
$U(\lambda N)=\lambda U(N)$. However, when long range interactions
are included, this property is violated, and it is easy to show that
thermodynamic functions are homogeneous of degree $1+|\alpha -1|$
for $\alpha<1$ (and logarithmic of $N$ for $\alpha=1$).

In addition, we expect that the internal energy of a magnetic system
with long-range interactions adopts the following form:
%\begin{equation}\label{Un}
   $U(S,M,N)=N{N}^*U_1(S/N,M/N),$
%\end{equation}
where $U_1$ is a function  per particle of the entropy,  $S$, and
magnetization, $M$. Generally, we could write other extensive
variables like $X=V,A,L,P,\cdot\cdot\cdot$ (representing the volume,
area, length, and polarization, respectively) as follows:
\begin{equation}\label{Ul}
   U(\lambda S,\lambda X,\lambda N)=
   \lambda^{1+|\alpha-1|}U(S,X,N).
\end{equation}
Hence, a problem with defining the thermodynamic limit persists for
$\alpha\leq 1$, in accordance with Eq.(\ref{Ul}).  The linearity is
only recovered for $\alpha>1$. As previously observed, a strong
dependence on the size of the system obstructs the behavior of the
thermodynamic functions and relations.
\subsection{Tsallis Conjecture:} As a possible way to solve this problem, Tsallis
conjectured that quantities like energy (like internal energy,
Helmholtz and Gibbs energies, and thermodynamic potentials per
particles) and intensive variables (like $T$ as in temperature, $H$
as in magnetic field) scale with $N^*$. Consistency of this
conjecture is shown as follows
\begin{equation}\label{GN}
   \frac{G_N}{N^*N}=\frac{U_N}{N^*N}-\frac{T}{N^*}\frac{S_N}{N}-\frac{H}{N^*}\frac{M}{N}
\end{equation}
in the thermodynamic limit. Observables per particle and intensive
variables are divergent when interactions are long-range. However,
scaling quantities are convergent anywhere. This kind of scaling
presents a standard thermodynamic structure because it preserves the
Euler and Gibbs-Duhem relations\cite{Abe}. By the same reasoning, it
is possible to define scaling quantities as follows:
$G_N^*=G_N/N^*$, $U_N^*=U_N/N^*$, $T^*=T/N^*$ and $H^*=H/N^*$. With
these definitions the previous equation is given by
\begin{equation}\label{GN}
   \frac{G_N^*}{N}=\frac{U_N^*}{N}-T^*\frac{S_N}{N}-H^*\frac{M}{N}.
\end{equation}
At this stage, we focus on some problems related to the
interpretation of the Eq.(\ref{GN}). For instance, the measured
temperature is not $T^*$, it is $T$; the measured internal energy
per particle is not $U_N^*$, it is $U_N$, etc.
\section{Periodic Boundary Conditions}
\label{PBC} If the range of microscopic interactions is smaller than
the size of the system, thermodynamic properties are obtained from
standard calculations. However, if interactions are long-range,
surface effects appear and begin to be important for all finite
sizes of systems. The calculation of thermodynamic quantities must
be carefully done. Long-range interactions are often defined as
those that do not fall faster than $1/r^D$, where $D$ is the space
dimension of the system. A first approach requires increasing
tremendously the size of the box. In general, periodic boundary
conditions are applied in order to eliminate surface effects in the
calculation of the thermodynamics properties of systems. This can be
exemplified by a central cell, which is repeated throughout space to
construct an infinite lattice. During the course of the present
study, if a particle moves in the central cell, its periodic images
move with the same orientation in everyone of the other cells. Thus,
as a particle leaves the central cell, one of its images will enter
through the opposite face. There are no walls at the boundary of the
central cell, and the system has no surface. In general, particles
interact with a potential of the following form $
 v(z) = \lim_{L \rightarrow \infty}\sum_{k=-L}^{L} g(|k + z|)
$, where the summation over $k$ represents all contributions over
replicated images. Thus, summations can be written in the following
manner:
 \begin{equation}\label{}
\sum_{k=-L}^{L} g(|k + z|)=\sum_{k=1}^{L} g(k -
z)+g(|z|)+\sum_{k=1}^{L} g(k + z).
\end{equation}
Two particular cases (the $1/z$ and the logarithmic potential) have
been discussed in a previous work\cite{PhysA340}.

Certainly, forces are obtained from $f(z) = -dv(z)/dz$. Due to the
symmetry of the lattice, it is easy to show that contributions to
the net force on every particle from all their images vanish. Thus,
periodic summations of forces do not practically depend on the
nature of interactions. In this manner, the net force
($f_N=-dg(z)/dz-F_L$) converges as quickly as the sum $F_L$
converges
\begin{equation}\label{sums}
F_L\sim\sum_{k=1}^{L} \left[\frac{dg(k + z)}{dz}+\frac{dg(k -
z)}{dz}\right],
\end{equation}
where $-1/2<z<1/2$ and $k\gg 1$. $dg(z)/dz$ represents the force of
a couple of particles in the central cell, and $F_L$ contributions
of all replicated cells in all space. If we take into account a
potential like Eq.(\ref{powerlaw}), net force converges in the same
manner as summations from Eq.(\ref{sums}):
\begin{eqnarray}
F_L\sim&&-\sum_{k=1}^{L} \left[\frac{1}{(k +
z)^{\alpha+1}}-\frac{1}{(k - z)^{\alpha+1}}\right] \nonumber
\\&&\approx 2 z (1+\alpha) \sum_{k=1}^{L}
\frac{1}{k^{2+\alpha}}\\&&\propto 4z \left( 1-L^{-1-\alpha}
\right).\nonumber\label{sums2}
\end{eqnarray}
If $L\geq 1$, contributions of particles and their images are
included.
\begin{figure}{}
\centering
%\vspace{-3cm}
\includegraphics[angle=270,width=1\textwidth]{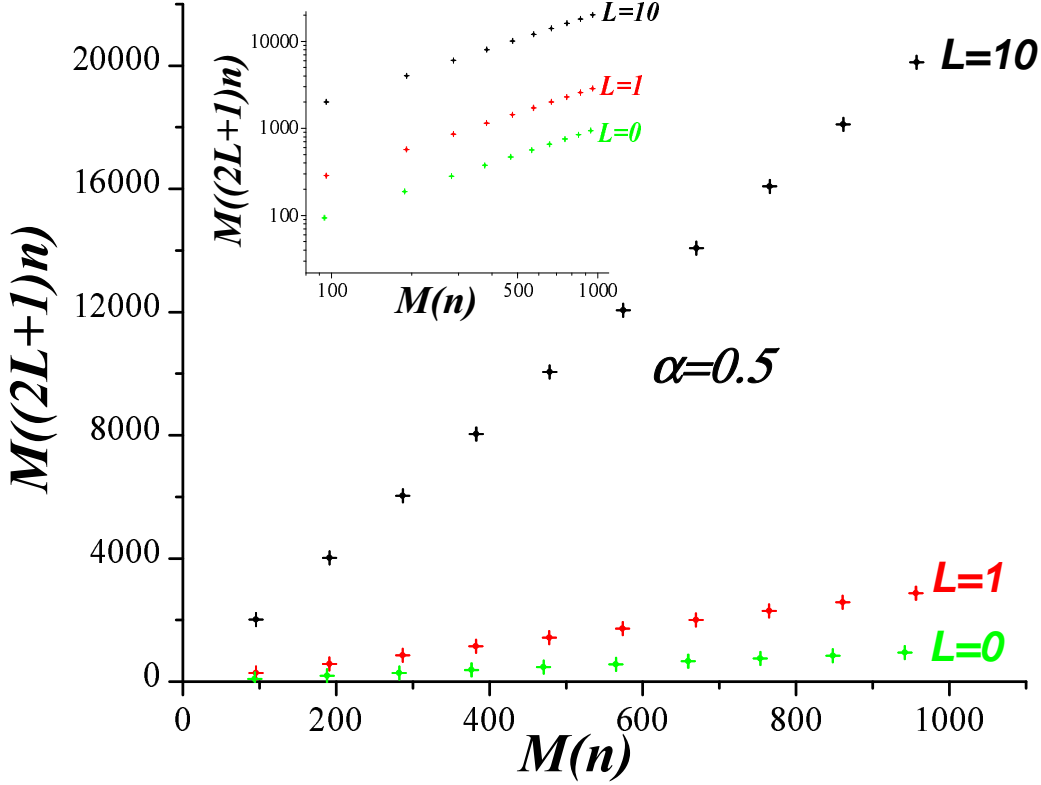}
%\vspace{-4cm}
\caption{Linearity of magnetization is shown for $L=0,\,1$ and $10$
and several values of $n$ with $N=(2L+1)n$. We plot $M(\lambda n)$
versus $M(n)$, where $\lambda=(2L+1)$. We see that the slope is
$\lambda$. In the inset we plot the same in log-log scale. The slope
is 1. This fact emphasizes that magnetization is a linear
homogeneous function of the size of the system.\label{linear} }
\end{figure}
\section{Scaling Hamiltonian and Numerical
Test}\label{scal} We consider a symmetric chain of $n$
spin-$\frac{1}{2}$ in a central cell that it is replicated $L$ times
over both sides. So, the total number of particles $N=(2L+1)n$.
Thus, we study the system via the Hamiltonian of the
Eq.(\ref{Hs})and apply periodic boundary conditions in the manner
introduced in section~\ref{PBC}. We write the coupling as follows
\begin{equation}\label{JK}
J(K)=\frac{1}{2}\sum_{l=-L}^{l=L}
\frac{J_\alpha^L(n)}{|nl+K|^{\alpha}}.
\end{equation}
Hence, in order to obtain  the scaling, we can determine it using
the tendency of internal energy, similar to Tsallis~\cite{Tsallis}.
Taking into account the symmetry of the chain and the continuous
limit, we derive the following integral:
\begin{eqnarray}\label{powerlaw2}
  U_N&\propto&2\int_{1}^{N/2} dr g(r)v(r) \nonumber \\
  &\propto&2\frac{(N/2)^{1-\alpha}-1}{1-\alpha},
\end{eqnarray}
where $g(r)$ (the pair distribution function) approaches 1 for $r\gg
1$. Indeed, from Eq.(\ref{powerlaw2}) the scaling coupling can be
written in the approach:
\begin{equation}
J_\alpha^L(n) = \left\{
\begin{array}{lll} J / 2^{\alpha}{N^{1-\alpha}} &\text{if $\alpha < 1$} \\
J / 2\ln N &\text{if $\alpha = 1$ }\\  ({1-\alpha}) J /{ 2}&\text{if
$\alpha > 1$ }
\end {array} \right. \label{Hs4}
\end{equation}
where $J_\alpha^L(n)$ measures the strength of the coupling that
depends on the size $n$ of the system. If we combine Eq.(\ref{JK})
and (\ref{Hs4}) and substitute the result into Eq.(\ref{Hs}), then
we obtain the scaling Hamiltonian that we used to carry out
numerical results with the Monte Carlo procedure. At this stage, we
emphasize that the scaling, in terms of $N^*$, is $2^\alpha{N^*}$
for $\alpha < 1$ and $2{N^*}$ for $\alpha \geq 1$. It has been well
expressed in Eq.(\ref{Hs4}) for symmetric lattices. Previously, (see
for instance Ref.\cite{prbCannas}) the scaling coupling was defined
in the old form $({1-\alpha}) J /{ 2^\alpha}$ and not as the new
form $({1-\alpha}) J /{ 2}$ (from Eq.(\ref{Hs4})) for $\alpha \geq
1$, for a discrete lattice. Hence, if the old scaling is applied to
systems whose most important geometrical property is symmetry, then
it would poorly represent the trends in thermodynamic quantities.

Thermodynamics describes the behavior of systems with many degrees
of freedom after they have reached a state of thermal equilibrium.
Furthermore, their thermodynamic state can be specified in terms of
a few parameters called state variables. At equilibrium, this method
of scaling the Hamiltonian allows observables to be linear with the
number of particles.\vspace{-0.5cm}
\subsection{Linearity of extensive quantities} It is well known that
thermodynamic quantities have to behave linearly relative to $N$,
the size of the system. For the present specific model of a chain of
spins with ferromagnetic long-range interactions, due to the form of
the scaling Hamiltonian, we expect that the linearity of internal
energy was completely satisfied and $U(\lambda{n})={\lambda}U(n)$.
This goes against the nonextensive view of the thermodynamics, which
predicts a nonlinear homogeneous behavior of internal energy, like
Eq.(\ref{Ul}). In addition, according to the Tsallis conjecture, we
expect that other extensive quantities also become linear with $N$.
In Fig.~\ref{linear}, the linearity of magnetization was tested from
simulations for $L=0,\,1$ and $10$ and several values of $n$, where
$N=(2L+1)n$. We depict several values of magnetization
$M(\lambda{n})$ versus $M(n)$, where $\lambda=(2L+1)$. A simple
inspection reveals linearity,
{\emph{i.e.}}$M(\lambda{n})={\lambda}M(n)$. Simulation points depict
right lines that differ only in their slopes. This fact is
highlighted in the inset of Fig.~\ref{linear} in log-log scale,
where we observe that all lines are parallel, whose slope is 1, an
indicator of linearity. So, we see that the linearity of an
observable is satisfied when it is different from the internal
energy, similar to the case of magnetization.
\subsection{Improvement of the critical temperature}
\begin{figure}{}
\centering
%\vspace{-3cm}
\includegraphics[angle=-90,width=1\textwidth]{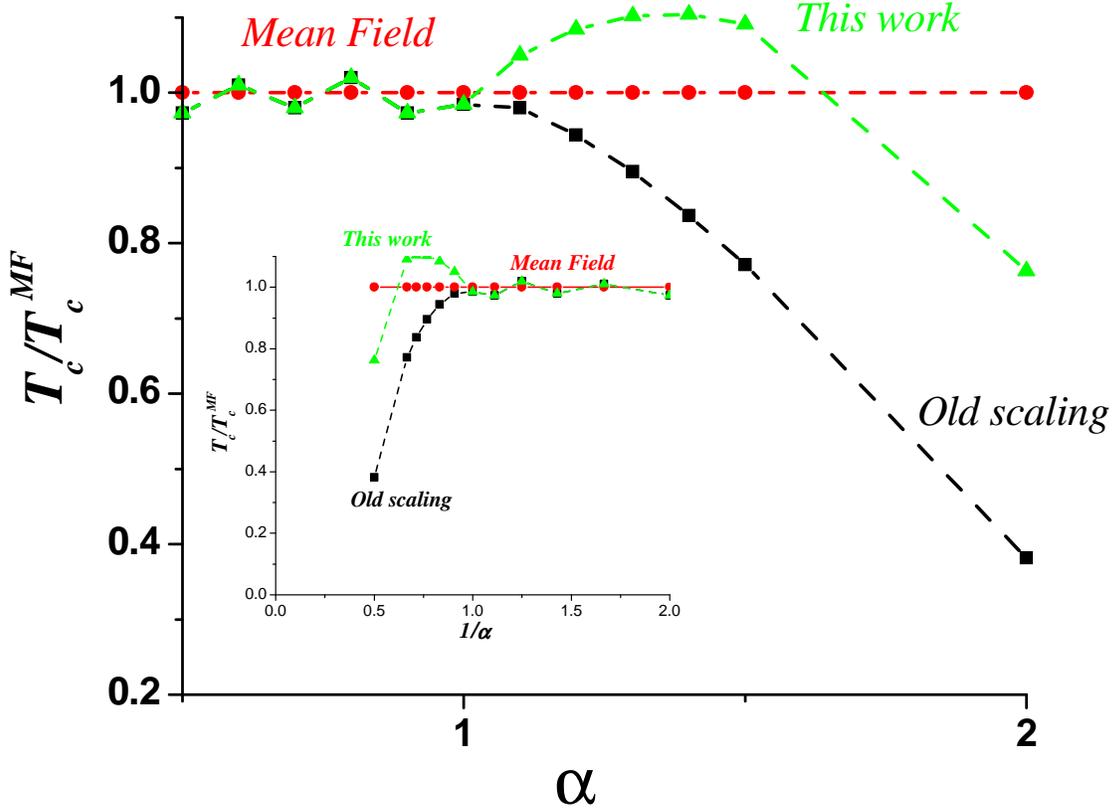}
%\vspace{-4cm}
\caption{The trend of the critical temperature as a function of
$\alpha$, the range of interaction. For $0\leq\alpha\leq 1$, the
critical temperature coincides with the mean field value. The new
trend is compared to the old one. In the inset we show the critical
temperature as a function of $1/\alpha$ \label{fig2} }
\end{figure}

Phase transition characterization is obtained in several ways. One
of the most adequate approaches is to define the critical point for
finite systems by the Binder method. To do so, we need to assess the
Binder cumulants of fourth order, which are defined as $u_n = 1 -
\langle{s}^4\rangle/3\langle{s}^2\rangle^2$. Cumulants $u_n$ as a
function of the temperature, intersect at a common point for several
sizes, $n$, of the system. This point is the critical temperature
that depends on the range of interactions $\alpha$. We test the
trend of the critical temperature and plot it as a function of
$\alpha$ using the scaling Hamiltonian, and we compare the present
result to the trend obtained previously from the old scaling. In the
present study, we carry out on a linear chain, where the number of
particles in the central cell is $10^2\leq n\leq10^3$. On the one
hand, the effective number of particles of the samples $N= (2L+1) n$
depends on the number, $L$, of replicas. On the other hand, the
computation time  depends just on $n$, not on $L$. We see that
$L=10^3$ is good enough, to compute in the thermodynamic limit and,
eliminate unnecessary sources of numerical fluctuations. For
$0\leq\alpha\leq 1$, the critical temperature approaches the mean
field value (\emph{e.g.,} for $\alpha=0.5$,
$T_c/T_c^{MF}=0.985\pm0.005$). But when $\alpha
> 1$, results from different scalings do not coincide and they are
different from the values predicted by the mean field approximation.
In Fig.~\ref{fig2}, the trend of the critical temperature is
depicted as a function of $\alpha$. We include results in the range
$0\leq\alpha\leq2$ for the critical temperature scaled in the manner
presented.
\section{Summary and Concluding Remarks}\label{conclu}
In the mean field approximation, the state of magnetic ordering of a
chain of spins with microscopic long-range interactions constitutes
an example of ferromagnetism in one dimension. Because interactions
are short range ({\em e.g.,} first neighbors) in the standard Ising
model, no magnetic ordering is observed in one dimension. These
approaches define two limiting cases. In the present work, the main
goal is to describe the crossover between these two limiting cases
in a suitable manner. First, if we take a scaling Hamiltonian, the
nice extensive property is recovered, and the thermodynamic
quantities are linear homogeneous functions against the homogeneous
function of degree $1+|\alpha +1|$, which are obtained if the
Hamiltonian is not scaled.
 Second, an improvement to the nonextensive scaling is proposed here
to obtain a more suitable method of describing the thermodynamic
behavior trend in this kind of systems for $\alpha > 1$. In the
$N\rightarrow\infty$ limit, we compare old scaling: $2^\alpha N^*$,
versus new scaling: $2 N^*$.
 Third, periodic boundary conditions can be used in the approach, using infinite replications of a central cell and considering the
 contribution over all space. It has been shown that forces converge rapidly. However, potentials increase with the size of the system.
The thermodynamic limit was reached in a numerical approach, with a
few particles in the central cell and a finite number, $L$, of
replications.
 Finally, we focus our attention on the Tsallis scaling (not the Tsallis conjecture); which is
 sufficient to solve the problem of the loss of linearity for
 thermodynamic quantities like energy and intensive variables, if we use it
 conveniently in the Hamiltonian.\vspace{-0.5cm}
\section*{Acknowledgments}
We would like to acknowledge partial financial support by FONDECYT
1051075 and 7060072.

\end{document}